# TEVisE: An Interactive Visual Analytics Tool to Explore Evolution of Keywords' Relations in Tweet Data

Shah Rukh Humayoun[1], Ibrahim Mansour[2], and Ragaad AlTarawneh[3]

[1] San Francisco State University, San Francisco, CA, USA
[2] University of Kaiserslautern, Kaiserslautern, Germany
[3] Intel Labs, Intel Corporation, Santa Clara, CA, USA
humayoun@sfsu.edu, ibrahim.mansour6@gmail.com, ragaad.altarawneh@intel.com

**Abstract.** Recently, a new window to explore tweet data has been opened in TExVis tool through visualizing the relations between the frequent keywords. However, timeline exploration of tweet data, not present in TExVis, could play a critical factor in understanding the changes in people's feedback and reaction over time. Targeting this, we present our visual analytics tool, called TEVisE. It uses an enhanced adjacency matrix diagram to overcome the cluttering problem in TExVis and visualizes the evolution of frequent keywords and the relations between these keywords over time. We conducted two user studies to find answers of our two formulated research questions. In the first user study, we focused on evaluating the used visualization layouts in both tools from the perspectives of common usability metrics and cognitive load theory. We found better accuracy in our TEVisE tool for tasks related to reading exploring relations between frequent keywords. In the second study, we collected users' feedback towards exploring the *summary* view and the new *timeline evolution* view inside TEVisE. In the second study, we collected users' feedback towards exploring the *summary* view and the new *timeline evolution* view inside TEVisE. We found that participants preferred both view, one to get overall glance while the other to get the trends changes over time.

**Keywords:** Information visualization, visual analytics, social media exploration, Twitter, tweet data, adjacency matrix diagram, Chord diagram, user study.

## 1 Introduction

Twitter has become a powerful tool for people to express their feedback, feelings, and reactions precisely towards recent topics. This create opportunities to researchers developing visual analytics (VA) tools to help users exploring and understanding underlying tweet data from different perspectives. For example, Nokia Internet Pulse [10] visualizes the evolution of Twitter discussions with a time series of stacked tag clouds, SparkClouds [18] integrates spark lines into the cloud tags to convey the trends between multiple tag clouds, and ScatterBlogs [34] visualizes geo-located Twitter messages.

We previously developed a tool, called TExVis [9], that enables exploration of data based on the relations between the frequent keywords. For this, TExVis used an enhanced Chord diagram to show the frequent keywords (e.g., hashtags, nouns, or verbs) and the relations between these keywords based on some criteria (e.g., the common tweets). In the TExVis user study [9], we showed that visualizing the relations between the frequent keywords based on some criteria helps users to explore and understand people' feedback and feelings towards an event. However, we found two main limitations of TExVis. First, it struggles from the visual cluttering issue in the relation area of the proposed extended Chord diagram due to the high number of relation chords between the frequent keywords (nodes), as mentioned by the participants in the TExVis user study. Second, TExVis shows the behavior of overall tweet data without considering the timeline evolution of underlying tweets. Such option can play an important role to enhance our understanding of how people's feedback and reaction are changed over time towards a particular event or product.

Targeting these limitations, we developed a visual analytics tool called **TEVisE** (**T**weets Timeline **E**volution **Vis**ual **E**xplorer). To avoid the visual cluttering, we decided to use an enhanced version of interactive adjacency matrix diagram to show the high-frequency keywords (e.g., hashtags, nouns, verbs) and the relations between these keywords based on certain criterion (e.g., the co-occurrence of keywords in same tweets). However, some researchers have suggested in the past that radial (circular) layouts produce compact visualizations and use the space efficiently than rectangular or square layouts [14, 15]. Keeping this in view, our *first research question* was which visualization layout, between the Chord diagram used in TExVis (i.e., a radial layout) and the adjacency matrix diagram used in TEVisE (i.e., a square layout), works better for the tasks related to reading and exploring the relations between the frequent keywords in tweet data.

To find out the answer of our first research question, we conducted a controlled user study with 12 participants in a lab environment. Our contributions in this regard are:

- A comparison of the visualization layouts in both tools to check the common usability aspects (i.e., *accuracy, efficiency,* and *user satisfaction*) with regard to exploring underlying tweet data. We were especially interested to see how the tasks related to the exploration of the relations between the frequent keywords effect the mentioned usability metrics.
- Assessing the cognitive load of the used visualization layouts in both tools to see their effect on users' limited working load memory capacity, which is utilized during information processing [26].

Targeting the second limitation in TExVis, we made enhancements in our TEVisE tool to show the *timeline evolution* of the frequent keywords and the relations between these keywords. With this new timeline evolution facility in TEVisE, our *second research question* was to analyze how users utilize the *timeline evolution* view for exploring an event compared to the overall *summary* view. To find the answer of this research question, we conducted a second user study with 10 new participants. Our contribution in this regard is:

- Getting the participants' subjective feedback to compare between the overall *summary* view and the *timeline evolution* view in order to see the usefulness of timeline view in exploring the underlying event.

The remainder of the paper is structured as follow. In Section 2, we provide the related work. In Section 3, we present our TEVisE tool with the enhanced adjacency matrix diagram and the timeline evolution view. In Section 4, we provide details of our first user study design and present the results. In Section 5, we describe our second user study design and present the results. In Section 6, we discuss the findings of both studies. Finally, we conclude the paper in Section 6 and shed light on future directions.

## 2 Related Work

As social media platforms (e.g., Twitter, Facebook) provide large amount of text data (e.g., tweets, comments, opinions, etc.), researchers developed powerful visualization and visual analytics tools to help users exploring and analyzing this huge amount of text data from different perspectives. Nokia Internet Pulse, developed by Kaye et al. [10], was one of the earliest visualization tools targeting tweet data. It visualizes the keywords frequency using a time-series of stacked clouds corresponding to a topic.

We see a deep interest in visualizing the sentiment analysis of tweets. One of the earliest works in this direction was done by Claster et at. [2], in which they visualized the sentimental polarity of over 80 millions tweets showing the effects on tourism in Thailand due to the unrest in early months of 2010. While Zhao et al. [40] examined Weibo, a Chines version of Twitter, provided timeline sentimental analysis of Weibo-based tweets in their MoodLens tool using four categories, i.e., *angry*, *disgusting*, *joyful*, and *sad*. Other examples are: Liu et at. [21] worked on the sentiment classification in tweet events. Nguyen et al. [29] used the case study of UK royal birth in 2013 and showed the sentimental polarity of tweets using a UK heat-map and a tile-map representation. Torkildson et al. [36] showed the sentiment frequency results of collected tweets, related to 2010 Gulf Oil Spill, through stacked-area chart. Wang et al. [37] developed SentiCompass to help users exploring and comparing the sentiments of time-varying tweet data and combined the 2D psychology model of motion with a time-tunnel representation. Kempter et al. [11] proposed multi-category emotion sentiment and visualized the summary of public emotion in Olympics 2012 related tweets. Lu et al. [22] used the map view and timeline chart view to show sentiments in tweet data of disaster scenarios. Munezero et al. [28] proposed the *enduring sentiments* concept that is based on psychological descriptions of sentiments, which is built over time while enduring emotional depositions. Mohammad et al. [27] combined the stance and sentiment analysis in tweets towards a target and showed the result using different visualizations (e.g., treemap, bar charts, etc.). Hoeber et al. [7] targeted sport event tweet data to visualize distribution of top terms, hashtags, user mentions, and authors in each of three sentiment polarity class using timeline charts geo-map. Dai and Prout [3] used a continuous word representation algorithm (Word2Vec) to train vector model on sentiment and collected Super Bowl 50 related tweets to demonstrate their classification method. Recently, Kucher et al. [16] developed StanceVis Prime tool for the analysis of sentiment and stance in temporal test data in social media data source.

There are also examples of tools that focus on analyzing and exploring tweet data based on the geo-spatial information. Few examples are: MacEachren et al. [23] developed SensePlace2 tool where they focused on using the geospatial information in tweets, obtained either through extracting information from the tweets or through user profiling location. Thom et al. used ScatterBlogs to visualize geo-located Twitter messages [34] and to study crisis intelligence [35]. Kraft et al. [12] focused on extracting the structural representations of events in tweet data. While Godwin et al. [6] developed TypoTweets Map tool for associating tweets to their spatial locations.

Exploring topics and analyzing their evolution over time has also been investigated by researchers. Few examples are in this directions are: Wanner et al. [38] developed Topic Tracker tool that uses shape-based time-series visualization to show trends and sentiment tracking of user defined topics. SparkClouds tool, developed by Lee et al. [18], integrates spark lines into the cloud tags to convey the trends between multiple tag clouds. Dork et al. [5] visualized tweet data in three modes: topics over time through Topics Streams layout, people and their activity through spiral layout, and popularity of event photos through Image Cloud. TopicFlow tool, developed by Malik et al. [24], visualizes the evolution of tweets for statistical topic modeling. Sopan et al. [32] did an analysis of academic conferences hashtags over time to analyze popular trends. Stojanovski et al. [31] visualized the topic distribution in the underlying tweet data in their TweetViz tool where users can also search for any hashtag or keywords. Recently, Li et at. [20] focused on the topic popularity in tweets over space and time.

Other researchers examined tweet data from different perspectives, e.g.: Krstajić et al. [13] focused on how real-time tweets can be used to early detecting any unexpected events. Kumamoto et al. [17] analyzed the impression building of a user through the posted tweets on the user's timeline. Humayoun et al. [9] developed TExVis tool to show the relations between the frequent keywords. Martins et al. [25] developed StanceXplore tool for the interactive exploration of stance in social media.

Except TExVis [9], other tools lack investigating the impact of relations between the frequent keywords inside tweet data. We tackle the limitations of TExVis tool from two directions: providing an enhanced adjacency matrix diagram to minimize the visual cluttering issue, and providing the timeline evolution of these relations through a series of proposed adjacency matrix diagrams.

## 3 TEVisE: Tweets Timeline Evolution Visual Explorer

TEVisE visualizes tweet data using an enhanced interactive adjacency matrix diagram (see Figure 1) to show the frequent keywords (e.g., hashtags, nouns, verbs) and the relations between these keywords based on a certain criterion (e.g., the *co-occurrence* of two keywords in same tweets). TEVisE was developed as a web-based tool using web technologies (e.g., HTML5, CSS, and JavaScript) and libraries such as *JQuery, d3.js* and *node.js* server.

As a proof of concept in this paper, we use the same dataset that was used by TExVis [9], which is consisted of 41,199 tweets (with 56,701 distinct keywords) and was extracted from Twitter using the "brexit" keyword. We assigned random ID numbers to individual tweets in order to anonymize the real users' IDs. The tweets (only in English and excluding the retweets) were fetches from Twitter using the *Tweetinvi* and *Twitter REST* APIs using a requesting loop for the time period between July 01 to July 10, 2016. The tweets were then tokenized based on hashtags, nouns and verbs using the *Apache OpenNLP* natural language processing library. Frequently used non-noun or non-verb hashtag were also separated as a token to use for frequent keywords. In order to classify tweets based on the contained sentimental polarities [30], we used the Aylien.TextApi library for each retrieved tweet. The library returns the polarity value (e.g., *positive*, *negative*, or *neutral*) alongside the confidence value (between 0% to 100%) of the stated polarity value. We used MongoDB as a database system to store the retrieved tweets and the above-mentioned data after pre-processing on stored tweets.

### 3.1 The Enhanced Adjacency Matrix Diagram

Some participants of TExVis user study [9] mentioned the visual cluttering problem in the relation area of the used Chord diagram (i.e., radial layout). Therefore, we decided to use a square layout (i.e., an adjacency matrix diagram) in TEVisE rather than a radial layout in order to avoid the visual cluttering issue. We argue that the implicit fashion of showing the relations using the intersections between the rows and columns in an adjacency matrix diagram reduces visual cluttering in the relation area. Due to this, we also argue that it enhances the readability of relations and supports users to preserves the mental map of the underlying data. Also, it is easy to compare a series of adjacency matrix diagrams in the case of timeline evolution.

We designed an enhanced interactive version of adjacency matrix diagram to show the high-frequency keywords, the relations between these keywords, and the sentimental polarity of the associated tweets to each keyword. Figure 1 shows the proposed enhanced adjacency matrix diagram using the "brexit" dataset having the most frequent twenty keywords. Initially, TEVisE shows only one adjacency matrix diagram, called the *summary* view, as in Figure 1, representing the overall summary result of all tweets collected against a specific keyword. This view can be changed into the *timeline evolution* view (see Figure 2(a)) to show the change over time, see the forthcoming section.

The nodes in our adjacency matrix diagram represent the high-frequency keywords (i.e., most popular *hashtags*, *nouns*, or *verbs*), which were extracted as tokens from the underlying tweet dataset. Initially, the nodes are arranged in the descending order based on keyword frequency, from top to bottom or left to right. A bar is attached to each node representing the corresponding keyword's frequency. The inside colors of these attached bars represent the division of sentimental polarity of associated tweets (i.e., green for *positive*, blue for *neutral*, and red for *negative*), where each color's length represents the frequency of associated tweets to a certain sentimental polarity. Mouse hovering over a particular node (keyword) or the attached bar shows further details through a tool tip (e.g., number of associated tweets, average confidence level, etc.).

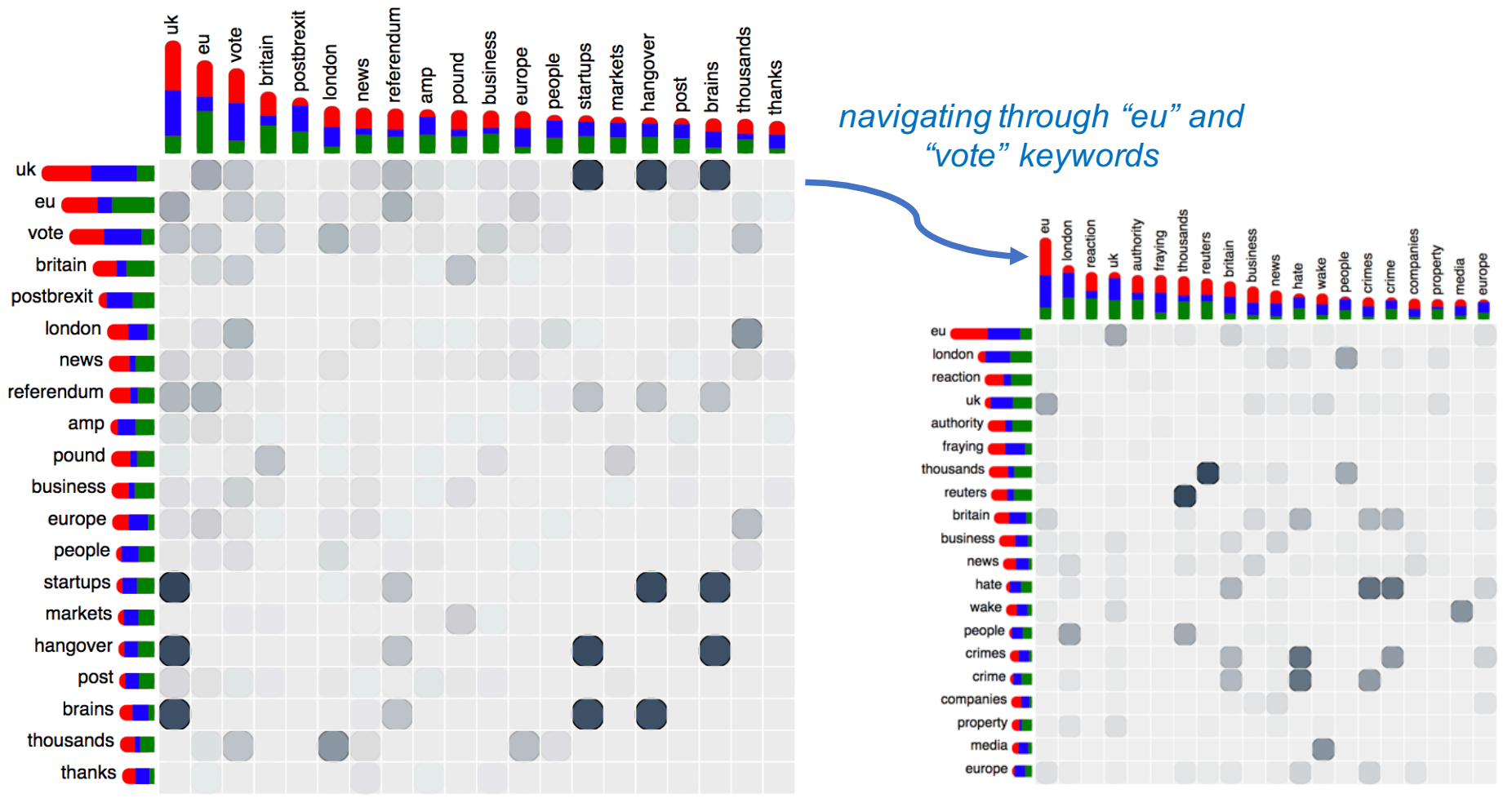

**Fig. 1.** The initial enhanced adjacency matrix diagram for "brexit" dataset *(left)* and the new diagram after navigating through "eu" and "vote" keywords (*right)*.

The matrix cells are used to show the relations between the frequent keywords, based on a certain criterion (e.g., in *co-occurrence* a relation occurs if both keywords are in the same tweet or the *word-similarity* relation of two keywords with a value of (0, 1] using the WordNet.Net library, etc.). The opacity of a matrix cell's grey color represents the value of the relation, e.g., Figure 1 (left) shows that people are talking more about "uk" and "startups" in same tweets, as these keywords show a high relation value in the co-occurrence relation. We decided to use only grey color scaling rather than color coding for the relation cells' values, so users would not confuse it with the attached bars' sentiment color coding. Mouse hovering over a particular matrix cell highlights this cell and the corresponding keywords in red color and brings a tooltip to show further details, e.g., the total number of associated tweets to this relation.

Navigation in the proposed enhanced adjacency matrix diagram is provided on-demand, which enables inspecting and exploring tweets that contain the selected keyword(s). This is achieved by clicking on a particular keyword or a matrix cell, which results in showing a menu with the option of navigation. In both cases, selecting navigation option results a new matrix diagram as a next level-of-details and deals with only those tweets that contain the selected keyword(s), see Figure 1. Selecting this option from a keyword brings the new matrix diagram with a one-level-down details, while selecting it from a matrix cell brings the new matrix diagram with a two-levels-down details. For example, navigating "eu" and then "vote" in two steps in Figure 1 brings a new matrix diagram related to only those tweets that have "brexit", "eu", and "vote" keywords together in them. This can be achieved directly in one step by navigating from the relation cell between the "eu" and "vote" keywords.

## 3.2 The Timeline Evolution of Tweets

Timeline analysis of tweets can play a critical role in enhancing our understanding of how people's feedback and reaction are changed over time towards a particular event or product. TEVisE uses a series of adjacency matrix diagrams (see Figure 2) to provide the timeline evolution of tweets, where each matrix diagram represents tweet data of a certain time period. In the *timeline evolution* view, the number of resulting matrix diagrams is based on the user's selection of time period over the total time period of underlying tweet data (e.g., based on hours, days, weeks, or months). These matrix diagrams are arranged from left to right and right to left at alternatively rows (starting from the top row) to provide a smooth viewing from one timeline to another one. TEVisE provides three categories to represent the timeline evolution of tweets:

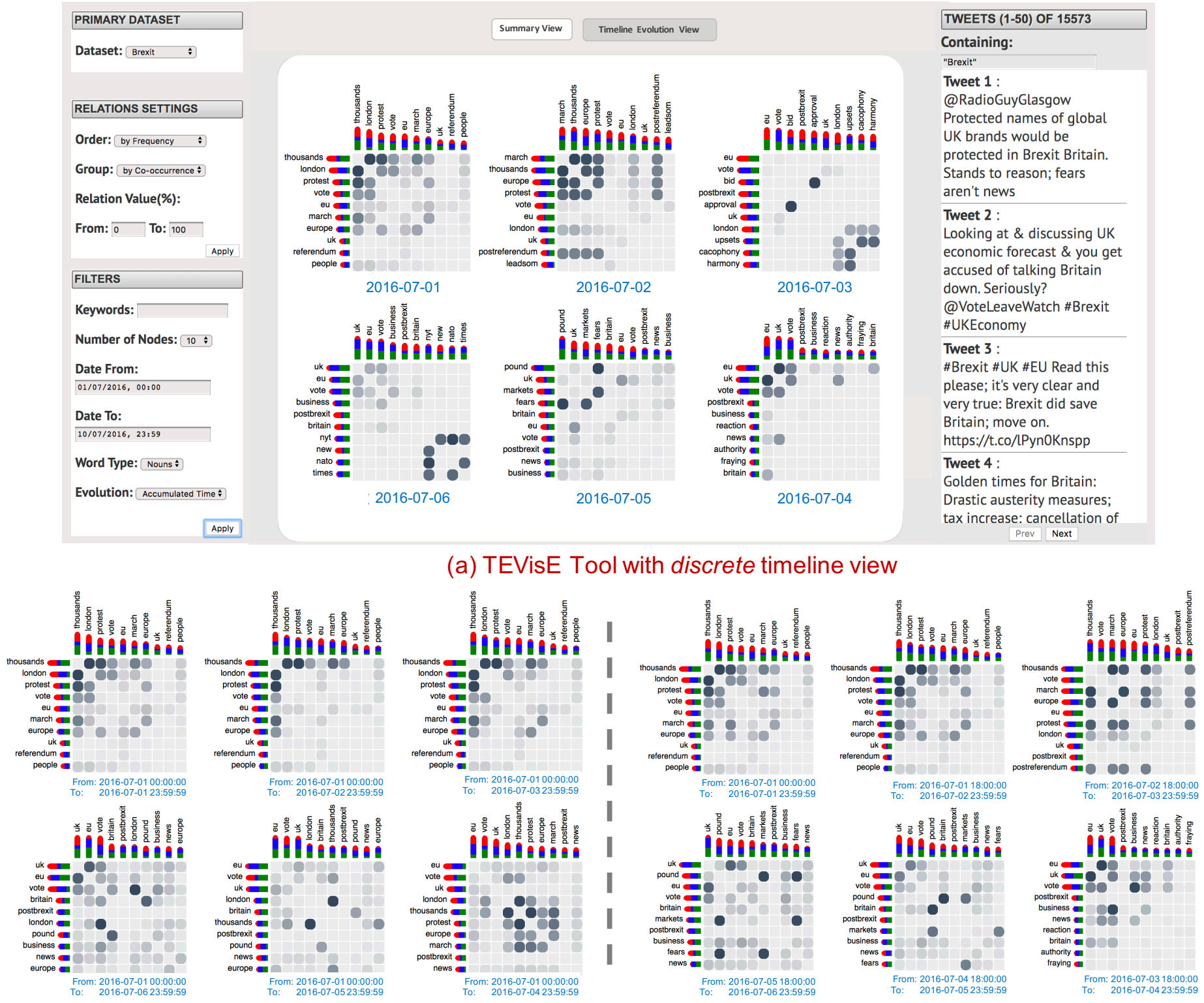

(a) TEVisE Tool with *discrete* timeline view

(b) *Accumulative* timeline view

(c) *Overlapping* timeline view

**Fig. 2.** TEVisE three categories of *timeline evolution* view (i.e., (a) *discrete*, (b) *accumulative*, and (c) *overlapping*) using the tweets from the first six days of "brexit" dataset.

- **Discrete Timeline Evolution:** In this case, each matrix diagram reflects tweet data of a certain time period in a way that the underlying data of this diagram is mutually exclusive with the data before that time period and after that time period. Figure 2(a) shows the “brexit” dataset in this view over the first six days of July 2016. In this case, each matrix diagram represents tweet data of only a particular day of the selected six days. Such discrete timeline visualizations are useful to explore the changes of people’s reaction and feedback from one time period to another one. For example in Figure 2(a), we see initially that on July 01 and 02 people talked more about those topics and keywords that are related to initial protests (e.g., “protest”, “march”, “london”, “thousand”); however, on July 04 and 05 people were more concerned about Europe future as they talked more about “eu”, “uk”, “vote”, and “post-brexit”. We also see the same trend in the relation cells between these keywords.

- **Accumulative Timeline Evolution:** In this case, each timeline division accumulates tweet data from the beginning to the current time period. Therefore, each matrix diagram reflects tweet data from the beginning to the current time period. For example, Figure 2(b) shows the “brexit” dataset over the first six days of July 2016, where the first matrix diagram reflects tweet data only of the first day, the second reflects tweet data of the first two days, the third reflects tweet data of the first three days, and so on. This is useful to explore the evolution of people’s reaction and feedback over time. For example in Figure 2(b), we see that the protest was initially the main topic; however, gradually the future of Europe and UK took place over it.
- **Overlapping Timeline Evolution:** This is a hybrid form between discrete and accumulative views. In this case, each matrix diagram reflects tweet data for a certain time period and accumulates it with the last quarter of the previous time period (except the first one). This is useful to show a smooth transition from one time period to another one. For example, Figure 2(c) shows the “brexit” dataset of the first six days of July 2016 in this style, where the second matrix diagram reflects tweet data starting from 18:00 of July 01 till the end of July 02, while the third matrix diagram reflects tweet data starting from 18:00 of July 02 till the end of July 03, and so on.

### 3.3 Interaction and Filtering Options

TEVisE provides a number of interactions and filtering options to help users better explore underlying tweet data. These interaction options are for both views, the *summary* view and the *timeline evolution* view. In the TEVisE tool, the original tweets associated to the current visualization(s) are shown in a tweet panel (see Figure 2(a)). These original tweets can be filtered out based on a specific keyword or a relation cell in the matrix diagram(s). Mouse hovering over a particular tweet in this panel also highlights all the associated keywords in the current matrix diagram(s).

A number of matrix sorting techniques have been proposed by researchers [1]. In TEVisE, we provide the option of sorting the nodes of matrix diagram(s) based on three criteria, i.e.: alphabetically by keywords names, keywords frequency (which is the default option) with descending or ascending order, or relation value. In the case of relation-based sorting, a keyword (node) gets the position (descending or ascending) based on the total of all corresponding relation values. In the *timeline evolution* view, mouse hovering over a keyword highlights this keyword and if presents then in all other matrix diagrams as well, while mouse hovering over a particular cell highlights this cell and the associated keywords and if presents then in all other matrix diagrams as well. Furthermore, individual zoom and panning facility is provided for all matrix diagrams.

In addition, users can filter the current matrix diagram(s) in both views through a number of filtering options using the filter-panel (see Figure 2(a)) in the tool, e.g.: selecting the relation type (e.g., *co-occurrence* or *word-similarity*), selecting relation percentage range (between any value from 0% to 100%), navigating the matrix diagram(s) based on a given keyword, selecting the number of nodes, selecting tweet data based on time by giving a starting and ending time value, and selecting what specific type of keywords should be shown (e.g., hashtags, nouns, verbs, or all). During the exploration and analysis of tweet data, users can switch between the summary view and the timeline evolution view while using all the interaction and filtering options, either provided directly on the matrix diagram or through the filtering panel.

## 4 User Study 1: TEVisE vs. TExVis

TExVis used radial layout (Chord diagram) [9], as it has also been claimed by researchers [14, 15] in the past that radial layouts encourage eye movement to proceed along the curved line of the circular diagrams, which helps viewers to better understand and explore the underlying data. However, we used the adjacency matrix diagram in TEVisE to avoid the visual cluttering in the relation area, which was mentioned by TExVis user study participants. In order to find the answer of our first research question, i.e., which layout amongst the used ones in both tools works better for the tasks related to reading and exploring the relations between the frequent keywords in tweet data, we conducted our first user study with 12 participants in a lab environment.

### 4.1 Study Goals

The goal of our first user study was to compare between the used visualization layouts in both tools from the perspective of common usability aspects, i.e., *effectiveness aka accuracy*, *efficiency*, and *user acceptance*. We were interested on how these usability aspects are effected due to the tasks related to reading and exploring the relations between frequent keywords using the visualizations in both tools.

Furthermore, we decided to investigate the cognitive load and the mental efforts that users needed while working with the underlying visualization layouts in both tools. Cognitive load theory was initially postulated by John Sweller [33] to measure the load of lessons and assessing learning. Current findings in the theory supports a three-factor approach, i.e., *intrinsic load:* the load associated with the material or subject matter, *extraneous load:* the load type predicated on the design, instructional and task complexity, and *germane load:* the load type measuring the elements contributing to learning [4]. In general, cognitive load theory is based on a limited working memory capacity, which is utilized during information processing [26]. Evaluating the used visualization layouts in both tools from the perspective of these three cognitive load factors would help us understanding the effect of the used visualization layouts on users' limited working memory capacity with respect to exploring tweet data.

### 4.2 Participants and Material

We were able to requite 12 participants (4 females and 8 males with average age of 27.3) on volunteer basis. The participants were either master-level students (i.e., 9) or recently graduated (i.e., 3). Amongst the participants, 10 were from computer science background while the remaining 2 were from electrical engineering background. All the participants reported not having any prior experience working with any of the underlying visualization layouts. Furthermore, none of them reported any color deficiency or any kind of color blindness.

The study was conducted as a *between-subject* study, where the participants were allocated randomly to either TEVisE or TExVis. Before starting an experiment, participants gave their informed consent and filled the pre-questionnaires form for collecting demographic information (i.e., age, gender, educational background, experience with underlying visualization, color deficiency or blindness). Then the participants were introduced to the experiment set-up with a 10-minute tutorial of the underlying tool and the used visualization. The same examiner was presented during all experiments. The study took place in a quite office environment using a 15.6-inch display laptop, where both web-based tools were installed locally to avoid any network latency. We used the "brexit" dataset as it was supported by both tools. We used only the *summary* view of TEVisE in this study, as TExVis does not support timeline evolution view.

Figure 3 shows two examples of the visualization layouts in both tools for this study. In the case of TExVis [9] (the Chord diagrams in Figure 3), nodes (arcs) represent the high-frequency keywords where a node's width represents the frequency, while the chords between nodes represent relations between the keywords based on some criteria (e.g., occurring in the same tweet). The width of a chord represents the relation value. Arcs outside of nodes are used to show the sentimental polarity of the associated tweets.

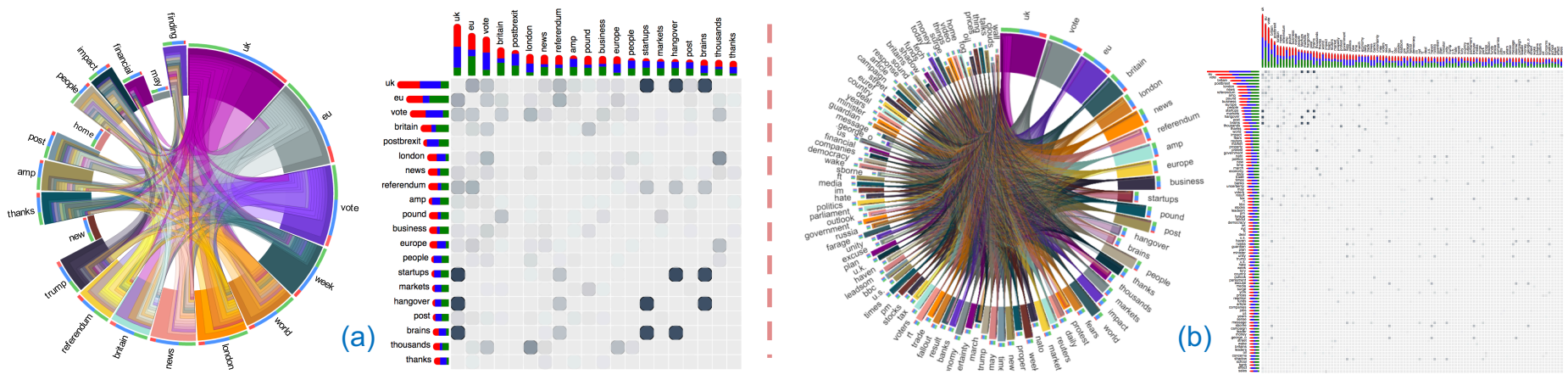

**Fig. 3.** The Chord diagram of TExVis [9] and the enhanced adjacency matrix diagram of TEVisE with 20 high-frequency keywords (a) and 100 high-frequency keywords (b). The participants in the study were allowed to configure the node size (frequent keywords) between 5 to 100.

### 4.3 Tasks, Hypotheses, and Metrics

We distilled 12 tasks in a way that encourages insight and sense-making and structured these 12 tasks into four categories (i.e., *overview*, *adjust*, *detect pattern*, and *match mental model*), according to the insight-gaining process by Yi et al. [39]. After the completion of each task category, participants were asked to fill out an adaptation version of the cognitive load questionnaire (with an eleven-point scale from 0 to 10) by Leppink et al. [19] to assess the cognitive load of the last tasks' category. There were 8 questions in this questionnaire classified into *intrinsic*, *extraneous*, and *germane* load sections. At the end, participants filled out a subjective closed-ended questionnaire with 10 questions using a 5-points Likert-scale, and an open-ended questionnaire.

We treated the used visualization as an independent variable, while the common usability aspects, i.e., effectiveness, efficiency, and user satisfaction, were taken as a dependent variables. Based on this, we formulated four hypotheses. Due to the visual cluttering issue in the relation area of the Chord diagram in TExVis:

- **H1:** We were expecting that TEVisE participants would achieve a higher accuracy value: *Accuracy(TEVisE) > Accuracy(TExVis)*
- **H2:** We were expecting that TEVisE participants would need less time to complete tasks compared to TExVis participants: *Time(TEVisE) < Time(TExVis)*
- **H3:** We were expecting that TEVisE would get higher user acceptance score in the subjective feedback than TExVis:
  *User Acceptance(TEVisE) > User Acceptance(TExVis)*
- **H4:** We were expecting that TEVisE would get lower rating of extraneous load (EL) and higher rating of germane load (GL):
  *EL(TEVisE) < EL(TExVis) & GL(TEVisE) > GL(TExVis)*

There were four tasks in the *overview* category with the nature of gaining insight about the underlying visualization, e.g., detecting nodes with specific qualities or detecting nodes having specific relation values; three tasks in the *adjust* category with the nature of filtering data and adjusting the underlying visualization, e.g., filtering the visualization per specific period of time or finding out nodes with a certain type of relation; three tasks in the *detect pattern* category with the nature of discovering detailed information from the underlying visualization, e.g., extracting some information from the visualization based on a certain keyword or identifying some relations with a certain values; and two comparative complex tasks in the *match mental model* category with the aim of ensuring how much the user has grasped the whole idea of underlying data as well as the underlying visualization, e.g., identifying and exploring keywords with higher positive or negative associated polarities.

### 4.4 Usability Results

Figure 4 shows the average accuracy per task for both tools. In the *overview* category, TEVisE achieved 93.75% overall average accuracy compared to 59.5% for TExVis. In the *adjust* category, TEVisE achieved 100% overall average accuracy compared to 63% for TExVis. In the *detect pattern* category, TEVisE achieved 83.33% overall average accuracy compared to 100% for TExVis. While in the *match mental model* category, both tools achieved 100% overall average accuracy. Overall, participants of TEVisE achieved 93.75% overall average accuracy in all twelve tasks compared to 77.25% for TExVis. When we check the results of both tools through applying the independent sample t-test statistically for hypothesis H1; we get $t = 2.02123$, $p\text{-value} = 0.027794$ at $p < 0.05$. These results support the acceptance of our hypothesis H1.

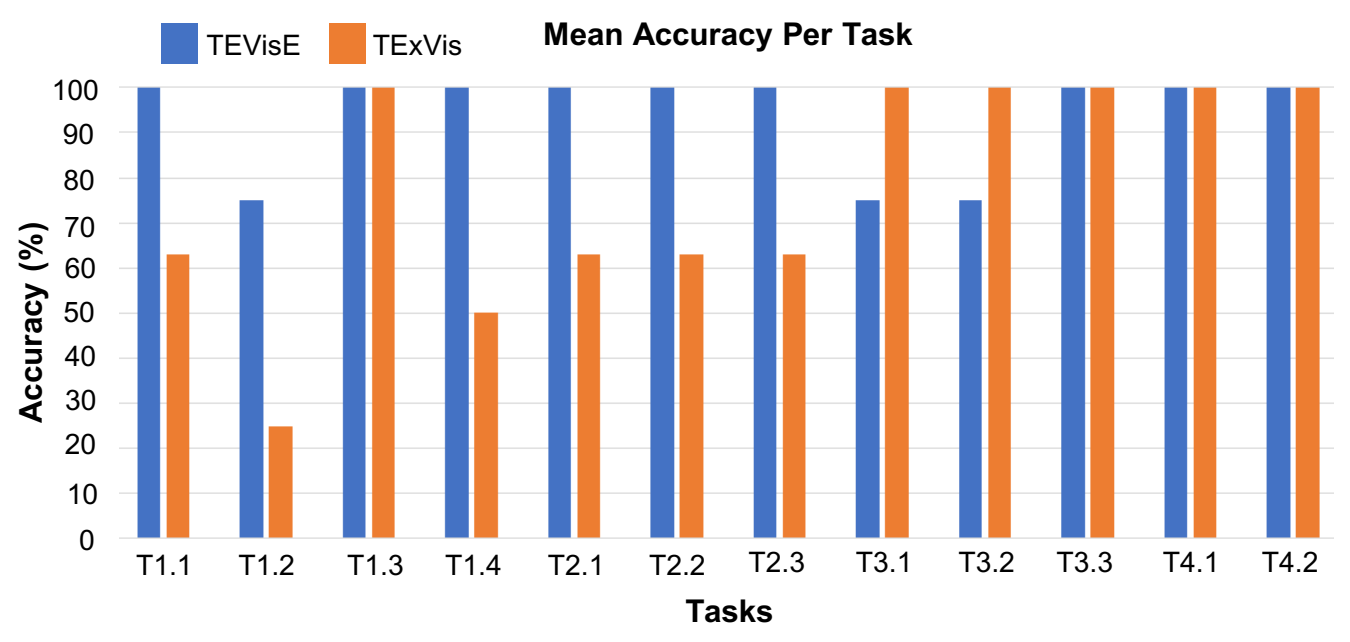

**Fig. 4.** Average Accuracy results per task for TEVisE and TExVis

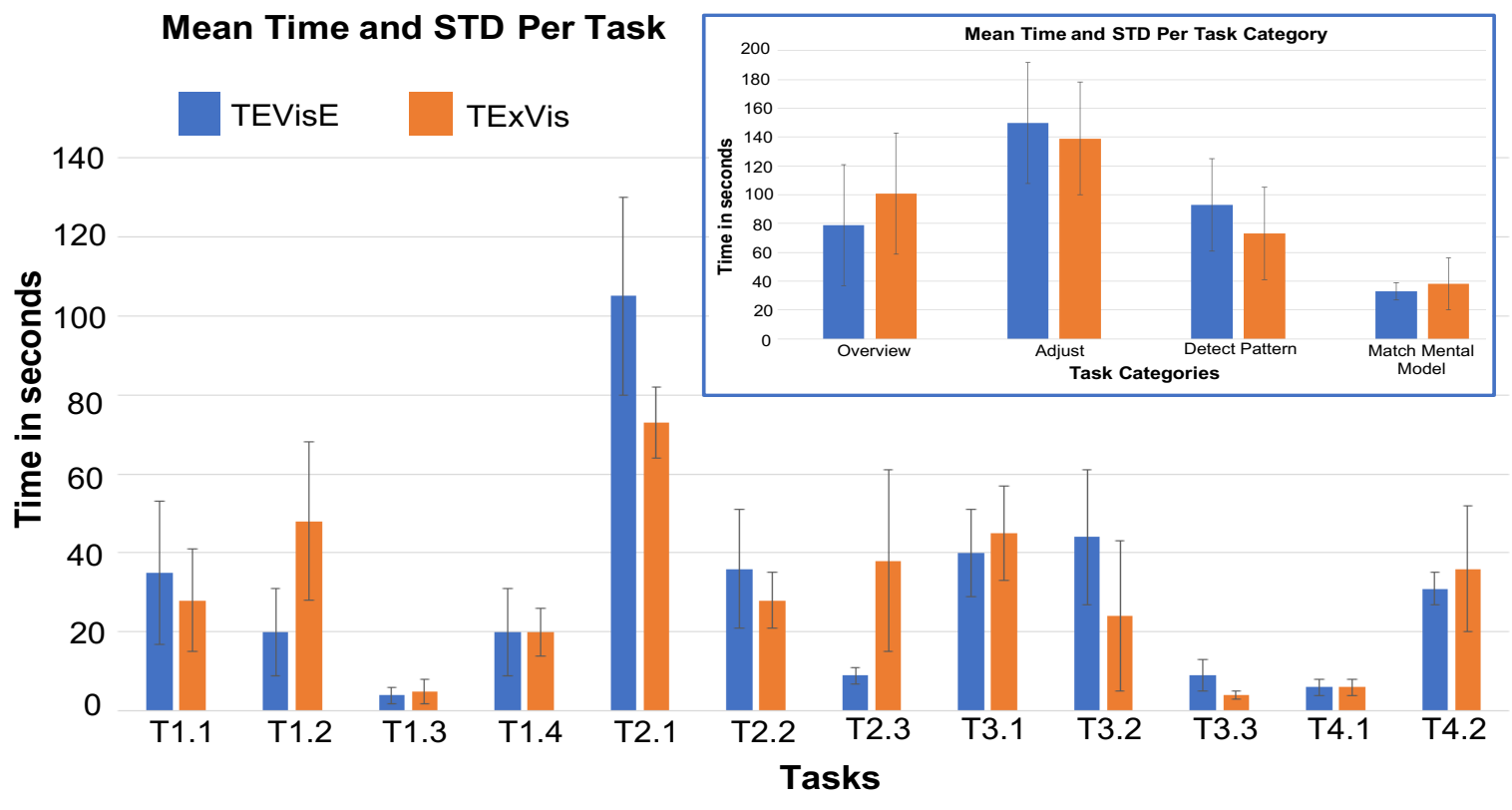

**Fig. 5.** Average efficiency results per task of TEVisE and TExVis participants. The top right-side box shows the average efficiency results per task-category.

We found out that TEVisE participants performed much better in those tasks that were related to identifying the relations between the keywords, which indicates the effectiveness of TEVisE in examining and exploring the relation area. Also, the cluttering appeared in the relation area of the Chord diagram played critical role for the low accuracy of TExVis participants in those tasks. On the other side, we found that TExVis participants performed better in those two tasks where they were asked to give answer related to a particular node. This can indicate that TExVis participants were more comfortable answering the tasks related to individual nodes, while TEVisE participants were more comfort for answering the tasks related to the relations among nodes.

Figure 5 shows the time required per task by participants of both tools and the overall time in each task-category. Participants using both tools achieved very comparative response with an overall average of 29.33 seconds for TEVisE compared to 29.58 seconds for TExVis. We found that participants had higher standard deviation values in the first two categories, which indicates that they reached to a common learning ground of the tool throughout the experiment duration. When we check the results of both tools through applying the independent sample t-test statistically for hypothesis H2; we get *t*

= *-0.02522*, *p-value = 0.490053* at $p < 0.05$. In this case, we failed to reject the null hypothesis. We think that participants spent more time in understanding the implicit relations in the TEVisE matrix diagram of. At the same time, we think due to the cluttering issue in the TExVis Chord diagram, participants' latency also got affected.

In the closed-ended questionnaire, TEVisE received higher average score on five questions, TExVis received higher average score on two questions, while both tools received same average score in the three questions. Overall, TEVisE received an average score of 4.2 compared to 3.9 for TExVis. When we check the results of both tools through applying the independent sample t-test statistically for hypothesis H3; we get $t = 1.19839$, *p-value = 0.123153* at $p < 0.05$. In this case as well, we failed to reject the null hypothesis. However, we found a major difference in the question 4, which was about the ease of discovering the relations between the keywords, where TEVisE received 4.5 score compared to 3.0 for TExVis. This again indicates the cluttering issue in the relation area of TExVis Chord diagram.

In the open-ended feedback, TExVis participants mentioned explicitly that they found it difficult to follow the relation-paths between the nodes due to the visual cluttering issue, even though filtering options were provided in the used Chord diagram. On the other side, TEVisE participants mentioned that the node-relation discovery in the used adjacency matrix diagram was useful in related tasks.

### 4.5 Cognitive Load Results

Based on the cognitive load theory's differentiation between the three loads, a minimum rating of extraneous load and a maximum rating of germane load is preferred in the results [19]. Intrinsic load is a load imposed by the inherent difficulty of the subject matter and cannot be influenced by the design of the visualization. Figure 6 shows the average factor scores for the three load types in all four categories of tasks.

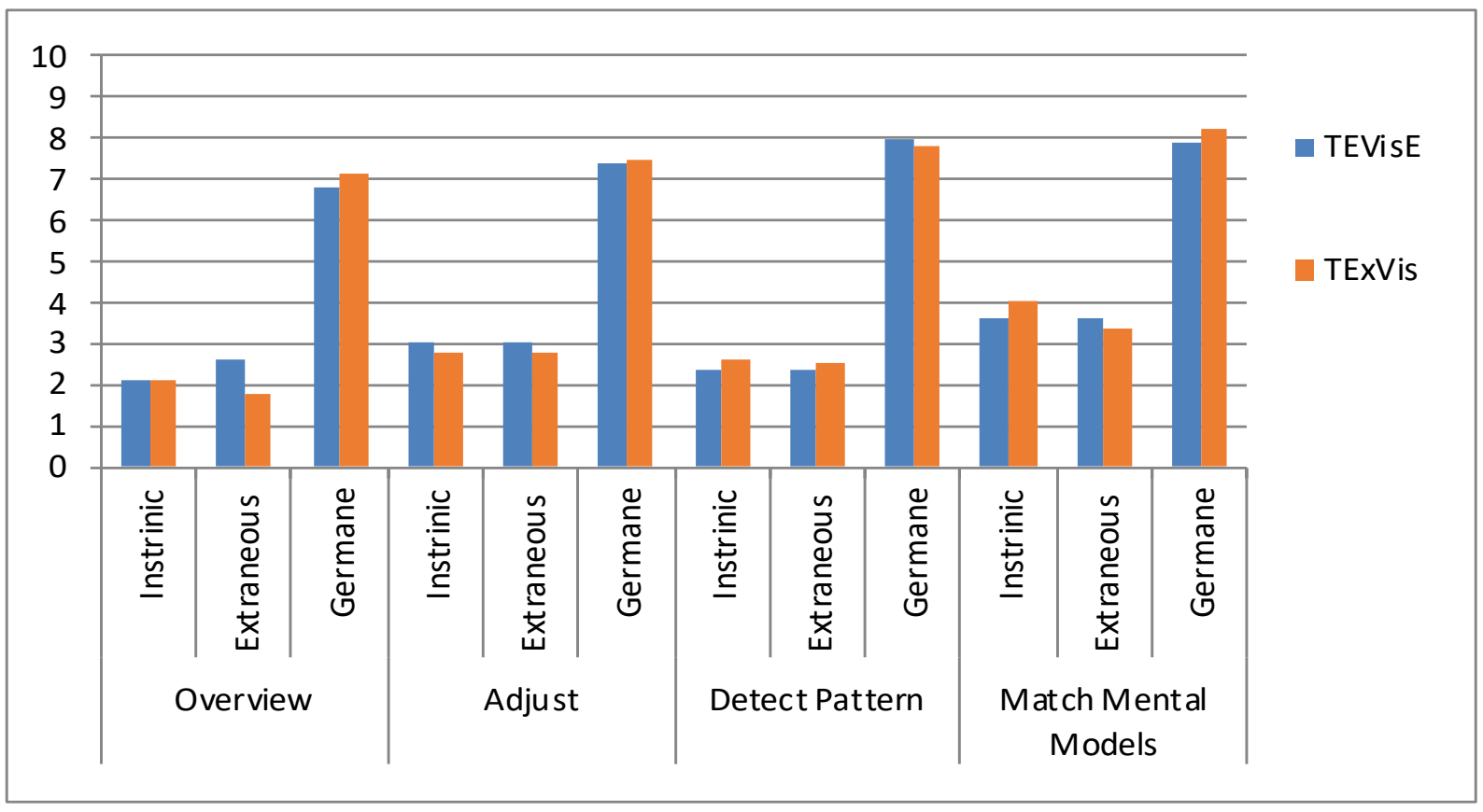

**Fig. 6.** Average factor scores for the three load types in the four categories of tasks.

As intrinsic load is independent of the resulting visualization; therefore, we were expecting the same kind of results in both tools. This is confirmed as the results show a quite similar score of intrinsic load in all categories of tasks between both tools. The extraneous load is based on the design, instructional and task complexity. In this case, the overall average score for TEVisE is 2.9 compared to 2.6 for TExVis ($t = 0.65311$, $p\text{-value} = 0.268945$ at $p < 0.05$). The difference between both tools is notably caused by the overview task category with an average extraneous load of 2.6 for TEVisE compared to 1.8 for TExVis. This could give an indication that the Chord diagram is more effective initially to provide the insight of underlying visualization at the beginning. The germane load is used to measure the elements contributing to learning. In this case, we do not find much difference between both groups, as TEVisE average rating is 7.5 compared to 7.6 for TExVis ($t = -0.26726$, $p\text{-value} = 0.399109$ at $p < 0.05$). In this case, we also failed to reject the null hypothesis.

# 5 Study 2: TEVisE Exploratory Study between the Summary View and the Timeline Evolution View

We conducted a second user study to find the answer of our *second research question*, i.e., to analyze how users utilize the new *timeline* evolution view for exploring tweet data for an event compared to the *summary* view. The study was conducted as a *within-subjects* style, where half of the randomly selected participants first used the summary view and then the timeline evolution view and the other half did it in the reverse order. Initially, we recruited 12 participants on volunteer basis; however, 2 participants were not able to work on both tools due to some technical issues; therefore, we excluded these 2 participants feedback form the study. The remaining 10 participants (3 females, aged 24-37, M = 28.4) were master-level students, where 9 of them were from computer science background and one was from business analytics background. We recruited those participants who did not participate in the first study in order to avoid any learning effects from the first study. Most of these participants did not have any prior experience of working with the underlying visualization layouts. Furthermore, none of these 10 participants reported any color deficiency or blindness.

Each experiment started with a 10-minute tutorial about the tool and the visualization layout, following a demographic questionnaire. We designed this study in an exploratory format; therefore, participants were asked to explore freely the underlying view for 10 minutes and analyze people's feedback and reaction towards the "brexit" event using the "brexit" dataset. They were free to use the filtering and sorting options provided by the tool, except to switch between the summary view and the timeline evolution view. During the experiment, participants were asked to take down notes of important findings. At the end of each view exploration, participants were asked to fill a closed-ended questionnaire with five questions using a 5 Likert-scale and an open-ended questionnaire with three questions to give their general feedback regarding the underlying view when analyzing the tweet data for exploring "brexit" event. At the end of both views' exploration, participants were also asked to provide their overall feedback about the both views. Each experiment lasted no more than one hour, including a 10-minute break between the two views' exploration. The study was conducted in a quite office environment using a 32-inches monitor, where the TEVisE tool was installed locally to avoid any network latency.

### 5.1 Results

In the first four questions of closed-ended questionnaire (i.e., *intuitiveness of the underlying visualization*, *easiness of finding the important information*, *support of underlying view to analyze the tweet data*, *and recommending the underlying view to use in future*), participants showed comparative more positive feedback towards the timeline evolution view compared to the summary view (i.e., 4 vs 3.7, 4.2 vs 4.3, 4.2 vs 3.9, and 4.3 vs 4.2 respectively for timeline evolution vs summary view). In the case of summary view, the fifth question was *"whether there is no need of timeline exploration view"* then all the participants, irrespective of whether they had already worked on the timeline evolution view or not, either strongly disagreed or disagreed (1.8 average score) with this statement. While in the timeline evolution view, the fifth question was *"importance of the timeline evolution exploration"*. In this case, most of the participants strongly agreed with the statement (4.8 average score). In spite of the same underlying visualization in both views, participants were better able to explore the underlying event and this can be the reason for getting a comparative positive feedback towards the timeline evolution view. The importance of timeline evolution view for better exploration is also shown from the feedback of question 5 in both cases.

In the open-ended questionnaire, we asked the participants to give their feedback about the important visual cues that they used in the underlying view when analyzing the "brexit" event. In the case of summary view, most of the participants told that they used all the important visual cues of the proposed matrix adjacency diagram like the frequency bars, the sentimental polarities, and the co-occurrence relation matric cells. However, few participants said that they focused more towards a particular visual cues among them. While in the case of timeline evolution view, most of the participants (i.e., 7) also mentioned explicitly the timeline in addition to the above-mentioned visual clues. They greatly appreciated the timeline evolution exploration of tweet data. Three examples are: *"Got a sense of time and how events were happening", "How easy it is to look at different combination of the data, and how the relationships change as you include some history"*, and *"I liked how easy it is to see trends change over a few days. The timeline view allowed me to see how the first couple of days were focused more on the protests, and how people started talking more about voting over time"*.

At the end of both views' exploration, we asked the participants to give their general feedback towards the both views. Many participants recommended that they would prefer to use both, e.g., one participant said: *"Both are necessary and complement each other"*. Participants mentioned that the summary view is good to get an overall glance of underlying tweet data, e.g., one participant said: *"Summary is useful to see the overall picture about an event. It can give us insights into the most talked topics and the relations between them"*. However, most of the participants mentioned that the timeline evolution view is important to see the changing of people's reaction and feedback over time, e.g.: *"The timeline mode can reveal more information about a particular day* *which can help users to dive deep into the data to explore the information"* or *"Timeline works better to see the changing trend"*. When asked which view is their first preference between both, 7 participants chose the timeline evolution view while 3 participants chose the summary view. These results indicate that both views help the user in exploring the event, the summary view for giving the accumulative glance of the event while the timeline evolution view for seeing the trends changes over time.

## 6 Discussion

We carried out two user studies to find the answers of our two research questions. Finding of our first user study can be summarized as the following:

- **Effectiveness aka accuracy:** In our first user study, TEVisE participants gained comparative higher accuracy compared to TExVis participants (i.e., overall average of 93.75% accuracy compared to 77.25% respectively), which indicates the effectiveness of the used enhanced adjacency matrix diagram in TEVisE. Mainly, TExVis achieved low accuracy in those tasks that were somehow related to the relations between the keywords, probably due to the visual cluttering issue in the relation area of the used Chord diagram. When we look into the results, we can infer that the used Chord diagram was useful when extracting information visualized on the curved line of the layout (e.g., information regarding nodes in it), as the radial layout encourages eye movement to proceed along the curved line of the underlying layout. However, the high density of relations between the nodes lowered down the accuracy due to the visual cluttering in the relation area of Chord diagram.
  Proper utilization of filtering and interaction options in the relation area of radial layouts can be useful for getting more accurate results. In TExVis, mouse hovering over a particular node results in fading of all other nodes' relations and changing the colors of this node's relations according to the opposite associated nodes. However, this did not help much TExVis participants in the related tasks, as keeping track of all or targeted relations was difficult due to visual cluttering. Therefore, we suggest investigating more about the kind of filtering and interaction options before applying them in the relation area of radial layouts. On the other side, the implicit fashion of using the transactions between the rows and columns in the used adjacency matrix diagram reduces the visual cluttering in the relation area, thus helped the TEVisE participants in executing the related tasks more accurately. Overall, we can infer that if tasks are more oriented towards extracting information regarding nodes, then radial layouts could be better choice for getting higher accuracy, while square layouts showing the relations implicitly (like our enhanced adjacency matrix diagram) could be better choice for tasks related to the relations between nodes.
- **Efficiency aka Latency:** In the case of time to complete the tasks, we received very mixed results, as in few tasks TEVisE participants performed better while in other tasks TExVis participants performed better. We also did not find any relation with efficiency to the accuracy. For example, in the tasks where TEVisE participants achieved higher accuracy, sometimes they took more time while in other cases they took less time than TExVis participants. In two tasks where TExVis participants achieved higher accuracy, we also see the same pattern. Overall both tools achieved the same latency average.

- **User acceptance:** The closed-ended feedback score endorses the accuracy finding, as TExVis received lower score in the relation related statement. This finding is further supported by the open-ended feedback, as some TExVis participants explicitly highlighted the difficulty of following paths to discover information about a relation due to the visual cluttering. As mentioned earlier that although the filtering options were provided in the used Chord diagram; however, it seems that TExVis participants became confused from the initial cluttering in the relation area of the diagram. On the other hand, TEVisE participants found the used matrix adjacency visualization useful in relation discovery and exploration. Few suggestions were provided by the participants for improving the interaction in our enhanced adjacency matrix diagram, e.g., highlighting the whole column and row when selecting a specific relation cell rather than just highlighting the corresponding keywords.
- **Cognitive Load:** Based on our results, TExVis showed a relatively better cognitive load indication rather than TEVisE. Although, a comparative low extraneous load rating and a slightly high germane load rating indicates the higher complexity of the adjacency matrix diagram compared to the Chord diagram; however, participants required nearly the same overall average time to finish the target tasks in both visualizations. We think that another reason of having low extraneous load score and high germane load score for TExVis might be due to the radial layout, as it encourages the eye movement to proceed along the curved lines rather than a zig-zag fashion in a square or rectangular figure [15]. Overall, participants of both visualizations mentioned an average of low mental effort for the *overview* and *adjust* sections, and very low mental effort for the *detect pattern* and *match mental model* sections.

With regard to answering our first research question, it is clear from the results that the adjacency matrix diagram used in TEVisE is more suitable for the tasks related to reading and exploring the relations between the keywords in tweet data, as it received higher accuracy score with nearly the same level of efficiency. Also, it provided a comparative overall higher user acceptance score with a clear difference in the question related to the relation area. On the other side, the Chord diagram used in TExVis showed a low extraneous load rating and a slightly high germane load rating, which indicates a comparative lower effect on users' limited working load memory when using this layout. However, the difference is not significantly different from TEVisE ratings. Therefore, we think it is comparable to use a square visualization layout, such as in TEVisE, in many real scenarios (e.g., analyzing tweets regarding an event for journalism, exploring tweets towards a new launched product to understand people's feedback, etc.), in order to get more accurate answers of relations-related questions. On the other side, we also suggest using the radial visualization layout, such as in TExVis, if emphasize is driven more towards tasks related to exploration of nodes. However, in such cases we suggest investigating more about the effectiveness of used filtering and interaction options before applying them in the relation area for getting higher accuracy.

There are some limitations in our study. Although, the used tweet data was real for the "brexit" event; however, the study was carried out with students only. Using a real interested user group for such event (e.g., journalists) would provide more insight of our first research question. Furthermore, we were able to compare only the summary view of TEVisE, as TExVis does not support timeline evolution view. Comparing the two visualizations from the timeline evolution perspective may reveal new findings both from usability and cognitive load perspectives, which is missing in our study. Also, the study was done with a limited pool of users; therefore, further studies are required in order to generalize the results between both visualization layout types.

In the case of second user study, the participants' feedback show the importance of timeline evolution view when exploring and analyzing people's feedback and reaction changing over time due to the continuous developments concerning the event. We also asked whether they would like to use both views or only one of the views, then 8 participants preferred to use both views while just two participants preferred to use the timeline evolution view only. This shows that both, the summary view as well as the timeline evolution view are necessary as one gives the overall glance of underlying tweet data while the other is useful to explore and analyze people's feedback and reaction over time for an in-depth analysis.

## 7 Conclusion and Future Work

In this paper, we presented our TEVisE tool to overcome the two main limitations of a previously developed tool TExVis, which struggles from the visual cluttering issue in the relation area and lacks in providing the timeline exploration of underlying tweet data. Visualizing the timeline of frequent keywords and their relations based on a certain criteria opens the doors to provide in-depth exploration of how people's reaction and feedback are changed over time towards some topics. Furthermore, such facility can be useful for many user groups and application domains, e.g., journalists can analyze tweets to explore and understand people's feedback and reaction over time towards an event, companies can analyze customers' reaction towards their new launched products, etc. We conducted two user studies to find the answers of two formulated research questions. We found out that using our enhanced adjacency matrix diagram in TEVisE increases the accuracy for the tasks related to reading and exploring relations between the frequent keywords. However, we found out that the Chord diagram used in the TExVis tool is slightly more preferable in terms of the cognitive load metrics. We discussed our general findings of both user studies in the previous section.

In the future, we intend to assess both views within our TEVisE tool (i.e., the summary view and the timeline evolution view) from the cognitive load perspective in order to see their impact on users' memory. We also want to analyze those factors that can improve the cognitive load in the case of cross-link multi-view visualizations such as the case of timeline evolution view. Furthermore, we plan to include additional functionalities in the tool, e.g.: the geo-spatial information in order to see people's feedback and reaction based on their locations (as few of the participants mentioned about it in their feedback), support of other social media platform (e.g., Facebook), and visual comparison of tweet data of two or more identical events/products in order to compare people's feedback and reaction towards these events/products.